\documentclass[prd,twocolumn,showpacs,floatfix,amsmath,nofootinbib,amssymb,floatfix]{revtex4}
\usepackage{graphicx,color,dcolumn,booktabs,bm,multirow}
\usepackage{longtable,lscape}
\usepackage{txfonts}
\usepackage{overpic}
\usepackage{amssymb}
\usepackage{indentfirst}
\usepackage{feynmf}   
\usepackage{slashed}  
\usepackage{cases}
\usepackage{color}
\usepackage{multirow}
\usepackage{epstopdf}
\usepackage{graphicx,color,dcolumn,booktabs,bm}
\usepackage{epstopdf}
\usepackage{ulem}

\usepackage[colorlinks,
            citecolor=green,
            anchorcolor=red,
            menucolor=red,
            linkcolor=red,
            filecolor=red,
            runcolor=red,
            urlcolor=blue,
            frenchlinks=red]{hyperref}

\newcommand{\sumint}{\kern 0.2 em {\textstyle\sum} \kern -1.1 em \int_X}

\begin{document}

\title{The electromagnetic form factors of $\Lambda$ hyperon in the vector meson dominance model}
\author{Yongliang Yang}\affiliation{School of Physics, Southeast University, Nanjing
211189, China}
\author{Dian-Yong Chen}\email{chendy@seu.edu.cn}\affiliation{School of Physics, Southeast University, Nanjing 211189, China}
\author{Zhun Lu}\email{zhunlu@seu.edu.cn}\affiliation{School of Physics, Southeast University, Nanjing 211189, China}

\begin{abstract}
We perform an analysis on the electromagnetic form factors of the $\Lambda$ hyperon in the time-like reaction $e^+e^-\rightarrow \Lambda\bar\Lambda$ by using a modified vector meson dominance model. We consider both the intrinsic structure components and the meson clouds components. For the latter one, we not only include the contributions from the $\phi$ and $\omega$ mesons, but also take into account the contributions from the resonance states $\omega(1420)$, $\omega(1650)$,  $\phi(1680)$ and $\phi(2170)$. We extract the model parameters by combined fit to the time-like effective form factor $|G_{\rm{eff}}|$, the electromagnetic form factor ratio $|G_E/G_M|$ and the relative phase $\Delta\Phi$ of the $\Lambda$ hyperon from the BaBar and BESIII Collaborations. We find that the vector meson dominance model can simultaneously describe these observables. Particularly, the inclusion of the resonance states in the model is necessary for explaining the ratio $|G_E/G_M|$ in a wide range of $\sqrt{s}$ as well as the large phase angle.
With the fitted parameters, we predict the single and double polarization observables, which could be measured in polarized annihilation reactions. Moreover, we analytically continue the expression of the form factors to space-like region and estimate the  space-like form factors of $\Lambda$ hyperon.
\end{abstract}

\maketitle
\section{introduction}

The electromagnetic form factors (EMFFs) $G_E$ and $G_M$ of hadrons are fundamental quantities for probing the internal structure of hadrons and understanding the perturbative and non-perturbative quantum chromodynamics (QCD) effects encoded in hadrons~\cite{Cabibbo:1960zza,Pacetti:2015iqa,Denig:2012by}.
They contain the information on the distribution of electric charge and magnetic moment of the hadron.
In the past decades the experimental and theoretical investigations on nucleon (proton and neutron) EMFFs ~\cite{Iachello:2004aq,Bijker:2005cd,Delcourt:1979ed,Bisello:1990rf,Armstrong:1992wq,
Bardin:1994am,Antonelli:1998fv,Kuraev:2011vq,Andreotti:2003bt,Brodsky:2003gs,Aubert:2005cb,
Pedlar:2005sj,TomasiGustafsson:2005kc,Lees:2013ebn,Achasov:2014ncd,Haidenbauer:2014kja,
Akhmetshin:2015ifg,Ablikim:2015vga,Pacetti:2015iqa,Denig:2012by}, particularly the effective form factor, have been performed in both the space-like and time-like regions, e.g., in the $ep$ elastic scattering, $\bar{p}p$ annihilation or $e^+e^-$ annihilation processes. A reasonable theoretical approach to understand the nucleon EMFFs in the space-like region is the vector meson dominance (VMD) model, which has been also extended to study the time-like data~\cite{Lomon:2002jx,Bijker:2005cd,Bijker:2004yu,Iachello:2004ki,Iachello:2004aq,Iachello:1972nu}.
In the framework of the VMD model, the EMFFs can be naturally expressed as the product of the two components: an intrinsic structure (the valance quark) and the meson cloud ($q\bar q$ pairs).
Other than the VMD model, the pQCD inspired model has also been applied to parameterize the nucleon EMFFs~\cite{Lepage:1979za,Lepage:1980fj,Belitsky:2002kj,Brodsky:2003gs,TomasiGustafsson:2005kc}, which have an analytical form that can also be extend to time-like region.

In recent years, there are also increasing interests on the EMFFs of the $\Lambda$ hyperon from both the theoretical~\cite{Haidenbauer:2016won,Haidenbauer:1991kt,Cao:2018kos,Yang:2017hao,Faldt:2017kgy,
Faldt:2016qee,Dalkarov:2009yf,Baldini:2007qg} and experimental side~\cite{Cui:2018,Bisello:1990rf,Aubert:2007uf,Ablikim:2017pyl}.
In contrast to nucleon, it is rather difficult to explore the scattering cross section and the EMFFs of the $\Lambda$ hyperon in space-like region~\cite{Ablikim:2017pyl,Dobbs:2014ifa,Aubert:2007uf,Bisello:1990rf}, since the hyperon are unstable and hyperon targets are unfeasible. The EMFFs of hyperons in the space-like region are hardly measured by exclusive experiments,
therefore, the time-like form factors in reaction $e^+e^-\rightarrow \Lambda\bar\Lambda$ can offer a unique opportunity to study the electromagnetic property of $\Lambda$ hyperon. Experimentally, the  cross sections for the $e^+ e^- \to \Lambda \bar{\Lambda}$ has been measured by BaBar and BESIII Collaborations \cite{Aubert:2007uf, Ablikim:2017tys,Ablikim:2017pyl}, a notable feature of the extracted effective form factor $G_{\rm{eff}}$ is the near threshold enhancement.  Similar enhancement effect can also be observed in the time-like effective form factor of proton~\cite{Aubert:2005cb,Lees:2013ebn,Akhmetshin:2015ifg,Achasov:2014ncd} and such phenomena can be interpreted by the Coulomb final-state interactions \cite{Lorenz:2015pba,Lichard:2018enc}, since the proton is a charged particle. As for $\Lambda$ hyperon, $\Lambda$ hyperon is neutral, thus, such kind of final-state interactions vanish. Then decoding the source of the near threshold of enhancement in the $\Lambda$ hyperon effective form factor will be a intrigue question.  Moreover, $G_E$ and $G_M$ in the time-like region are complex, there is a relative phase angle $\Delta \Phi$ between these two form factors. Very recently, a preliminary measurement~\cite{Cui:2018} from the BESIII Collaboration demonstrates a rather large phase $\Delta \Phi=42^\circ \pm 16^\circ \pm 8^\circ$ at $\sqrt{s}=2.396\,\rm{GeV}$.
To date,  there is very little theoretical implication on the relative phase $\Delta \Phi$ other than the $\Lambda \Lambda$ potential model ~\cite{Haidenbauer:2016won}.

In the present work, we aim at simultaneously describing the available data on the effective form factor $|G_{\rm{eff}}|$, the ratio $|G_E/G_M|$ and the relative phase $\Delta\Phi$ (preliminary data) from both BaBar and BESIII Collaborations, in light of the VMD model. Due to the isoscalar property of the $\Lambda$, we need not consider the contribution of the $\rho$ meson and its resonance states. Furthermore, the threshold of $\Lambda \bar{\Lambda}$ is about 2231 MeV, thus $\omega$ and $\phi$, as well as their excited states below the threshold should be involved. In the present work, we take into account the contributions from the resonance states $\omega(1420)$, $\omega(1650)$, $\phi(1680)$ and $\phi(2170)$.  The formula of time-like form factors are obtained by an analytic continuation of the space-like form factors, and in the time-like region, we take into account the decay widths of the vector mesons and their resonance states in order to introduce a complex structure for $G_E$ and $G_M$~\cite{Lorenz:2015pba}.

The work is organized as follows: an analysis of the form factors of $\Lambda $ hyperon in the VMD model is performed after introduction. In section \ref{Sec:3}, we present our fit to the time-like form factor of $\Lambda$ hyperon and our predictions for the single and double polarization observables as well as space-like form factors and Section \ref{Sec:4} is devoted to a short summary.

\section{Analysis of form factors of $\Lambda$ hyperon in the VMD model \label{Sec:2}}

Encouraged by the success of the VMD model for the nucleon (proton and neutron) in Ref.~\cite{Iachello:1972nu},
we extend the model to investigate the EMFFs of the $\Lambda$ hyperon. First, we introduce the $\Lambda$ form factors in the space-like region, where $q^2=-Q^2<0$.
Considering the relativistic invariance of the EMFFs, one can write down the electromagnetic current of a  baryon with spin-$1/2$ in terms of the Dirac form factor $F_1(Q^2)$ and Pauli form factor $F_2(Q^2)$ as
\begin{align}
J^\mu=\gamma^\mu F_1(Q^2)+{i\sigma^{\mu\nu}q_\nu\over 2M_\Lambda}F_2(Q^2)\, .
\end{align}
In the VMD model,  $F_i$ can be further decomposed as
\begin{align}
F_i = F_i^S + F_i^V
\end{align}
where $F_i^S$ and $F_i^V$ denote the isoscalar and isovector components of the form factors, respectively. As for $\Lambda$ hyperon, there is only isoscalar contribution, i.e., $F_i^V \equiv 0$, since $\Lambda$ hyperon is isospin singlet.

Based on the above consideration, the observed EMFFs $G_E(Q^2)$, $G_M(Q^2)$ of the $\Lambda$ hyperon can be expressed in terms of the $F_i^S$ as,
\begin{align}
G_{M_\Lambda}=F_1^S+F_2^S,~~~~~G_{E_\Lambda}=F_1^S-\tau F_2^S\,,
\label{eq:gegm}
\end{align}
where $\tau={Q^2/4M^2_\Lambda}$. We note that these relations satisfy the kinematical constraint $G_E(-4M_\Lambda^2)=G_M(-4M_\Lambda^2)$.
This constraint is of crucial importance in the time-like region~\cite{Iachello:2004ki}.

In the VMD models for the nucleon~\cite{Bijker:2005cd,Bijker:2004yu,Iachello:2004aq,Iachello:1972nu}, the form factors are attributed to two parts. One is the intrinsic structure which is determined by the valence quarks, the other is the meson cloud.
The nature of the intrinsic structure has been analyzed and discussed in details in Refs.~\cite{Bijker:2005cd,Bijker:2004yu}. In those studies the form factors satisfy the asymptotic behaviour of pQCD~\cite{Belitsky:2002kj,Brodsky:2003pw}.
The meson cloud term was used to describe the interaction between the bare nucleon and the photon in the framework of the vector meson dominance ($\rho$, $\omega$ and $\phi$)~\cite{Iachello:2004ki,Iachello:2004aq}.

\subsection{space-like form factors}

In light of the VMD, we consider the isoscalar components of the form factors which receive contributions from the $\omega$ and $\phi$ mesons.
The Dirac and Pauli form factors of the $\Lambda$ hyperon are expressed as the product of the intrinsic form factor $g(Q^2)$ and the terms from VMD.
In addition, we will also consider the contributions from the resonance states of the vector meson $\phi$ and $\omega$: $\omega(1420)$, $\omega(1650)$, $\phi(1680)$ and $\phi(2170)$.
The expressions of the form factors from these resonance states are assumed to be the same as those from the vector meson $\omega(782)$ and $\phi(1020)$ in our modified model. The Dirac and Pauli form factors of $\Lambda$ hyperon are normalized at $Q^2=0$ as
\begin{align}
F_1(0)=0,~~~~~ F_2=\mu_\Lambda\,.
\end{align}
These constraints lead to the parameterized forms of the isoscalar Dirac and Pauli form factors in the VMD model as follow,
\begin{eqnarray}
F_1^S(Q^2)&=&{g(Q^2)\over 3}\Sigma_{i=1}^N\bigg{[}-\beta_{\omega_i}-\beta_{\phi_i}+\beta_{\omega_i}
{m^2_{\omega_i}\over m^2_{\omega_i}+Q^2} \nonumber\\ &&+\beta_{\phi_i}{m^2_{\phi_i}\over m^2_{\phi_i}+Q^2}\bigg{]}\,\label{F1s}\\
F_2^S(Q^2)&=&{g(Q^2)\over 3}\Sigma_{i=1}^N\bigg{[}(\mu_{\Lambda}-\alpha_{\phi_i}){m^2_{\omega_i}\over m^2_{\omega_i}+Q^2}+\alpha_{\phi_i}{m^2_{\phi_i}\over m^2_{\phi_i}+Q^2}\bigg{]}\,\nonumber \\
\label{F2s}
\end{eqnarray}
where $N=3$ and $\mu_{\Lambda}=-0.613\mu_N$ is the magnetic moment of the $\Lambda$ hyperon,
$\omega_i \ (i=1,2,3)$ denotes the vector meson states $\omega(782)$, $\omega(1420)$ and $\omega(1650)$, and $\phi_i\ (i=1,2,3)$ denotes the vector meson states $\phi(1020)$, $\phi(1680)$ and $\phi(2170)$.
The intrinsic structure factor is characterized by a three-valence-quark structure~\cite{Bijker:2005cd}. We choose it as a dipole form $g(Q^2)=(1+\gamma Q^2)^{-2}$, which is consistent with pQCD and enables a good fit to the nucleon EMFFs~\cite{Bijker:2004yu,Iachello:1972nu}.
Especially, we adopt the value of the intrinsic parameter $\gamma=0.25$ as the one in Ref.~\cite{Iachello:1972nu}.
While, the coefficients $\beta_{\omega_i}$, $\beta_{\phi_i}$, $\alpha_{\phi_i}$ in Eqs.~(\ref{F1s})-(\ref{F2s}), which are products of a $V\gamma$ coupling constant and $VBB$ coupling constant, are considered as free parameters~\cite{Iachello:1972nu}. Since our purpose is to investigate the existing data for time-like form factors, we do not further elaborate the space-like expressions. We will extend the expressions of the form factor to the time-like region in next subsection.

\subsection{Time-like form factors}

The general expression of the Born cross section for the reaction $e^+e^-\rightarrow \bar B\,B$ has been given in Ref.~\cite{Haidenbauer:2014kja}
under the one-photon exchange approximation with $B$ as a spin-$1\over 2$ baryon.
The integrated cross section of the $\Lambda$ hyperon pairs production is governed by the electric and magnetic form factors $G_E$ and $G_M$ as,
\begin{align}
\sigma(s)={4\pi\alpha^2\beta\over 3s}C_\Lambda\bigg{[}|G_M(s)|^2+{1\over 2\tau}|G_E(s)|^2\bigg{]}\,
 \label{eq:crosssection}
\end{align}
where $G_E(s)$ and $G_M(s)$ are the EMFFs in the time-like region, $s$ is the square of the center of mass (c.m.) energy, $\tau={s/ 4M^2_\Lambda}$.
The variable $\beta=\sqrt{1-1/\tau}$ is a phase-space factor.
The Coulomb enhancement factor $C_\Lambda$~\cite{Baldini:2007qg,Arbuzov:2011ff}, accounting for the electromagnetic interaction of the point-like baryon pairs in the final state, is $C_\Lambda=1$ for neutral baryon pair.
Another quantity used in various analyses is the effective form factor $G_{\text{eff}}(s)$.
The effective form factor is equivalent to $|G_M|$ under the hypothesis $G_E=G_M$~\cite{TomasiGustafsson:2005kc}.
In general cases, the effective form factor can be expressed in terms of the moduli of the EMFFs~\cite{Aubert:2005cb}:
\begin{align}
|G_{\rm{eff}}(s)|=\sqrt{2\tau|G_M(s)|^2+|G_E(s)|^2\over 1+2\tau}=\sqrt{\sigma_{e^+e^-\rightarrow \bar\Lambda\Lambda}(s)\over {4\pi\alpha^2\beta\over 3s}C_\Lambda[1+{1\over 2\tau}]}\,.
\label{eq:Geff}
\end{align}
It is proportional to the square root of the Born cross section, which can be extracted from experimental measurements on the cross section for $e^+e^-\to \Lambda \bar{\Lambda}$.

By an appropriate analytic continuation in the complex plane, the form factors of Eqs.~(\ref{F1s})-(\ref{F2s}) can be used to analyze the form factors in the time-like region~\cite{TomasiGustafsson:2005kc,Iachello:2004aq}. For the intrinsic structure, the analytical continuation to time-like region is based on the following relation~\cite{TomasiGustafsson:2005kc}:
\begin{align}
Q^2= -q^2=q^2 e^{i\pi}\,.
\end{align}
Therefore, in the time-like region, $g(q^2)$ has the form an analytical continuation form:
\begin{align}
 g(q^2)={1\over (1-\gamma q^2)^2}.
\end{align}
For the meson cloud part, we take into account the width of vector mesons~\cite{Lorenz:2015pba} to introduce the complex structure in the time-like region. By the following replacement,
\begin{eqnarray}
 \beta_{\omega_i}{m^2_{\omega_i}\over m^2_{\omega_i}+Q^2} &\rightarrow& \beta_{\omega_i}{m^2_{\omega_i}\over m^2_{\omega_i}-q^2-i m_{\omega_i}\Gamma_{\omega_i}}\,, \nonumber \\
 \beta_{\phi_i}{m^2_{\phi_i}\over m^2_{\phi_i}+Q^2}&\rightarrow& \beta_{\phi_i}{m^2_{\phi_i}\over m^2_{\phi_i}-q^2-i m_{\phi_i}\Gamma_{\phi_i}}\, ,
 \label{Eq:Rep}
\end{eqnarray}
we can obtain the modified VMD model in the time-like region. It should be noticed that in the case of nucleon, the time-like form factors have been analyzed in a VMD model ~\cite{Bijker:2005cd,Iachello:2004aq}, where only the ground vector mesons were included. It was found that the complex structure of nucleon time-like form factors came from a undetermined phase angle $\theta$ in $g(q^2)$ as well as the width of $\rho$ meson \cite{Frazer:1960zzb}. In the present model, we include the excited states of the vector meson states and the widths of involved states are crucial for reproducing the relative phase angle $\Delta \Phi$, which is similar to the case of the pion form factor studied in Ref.~\cite{OConnell:1995nse} through the VMD model.

\begin{table}[htb]
\caption{The masses and widths of the involved vector mesons in unit of MeV \cite{Tanabashi:2018oca}. \label{Tab:Mass}}
\begin{tabular}{p{1.2cm}<{\centering}p{1.2cm}<{\centering}p{1.2cm}<{\centering}p{1.2cm}<{\centering}p{1.2cm}<{\centering}p{1.2cm}<{\centering}}
\toprule[1pt]
State & Mass & Width &State &Mass & Width \\
\midrule[1pt]
$\omega(782)$  &783 & 8.5 & $\phi(1020)$ &1019 & 4.2 \\
$\omega(1420)$ &1420 &220 & $\phi(1680)$ &1680 & 150 \\
$\omega(1650)$ &1650 &315  &$\phi(2170)$ & 2188 & 83\\
\bottomrule[1pt]
\end{tabular}
\end{table}

\begin{figure}[htb]
  \centering
  \includegraphics[width=0.95\columnwidth]{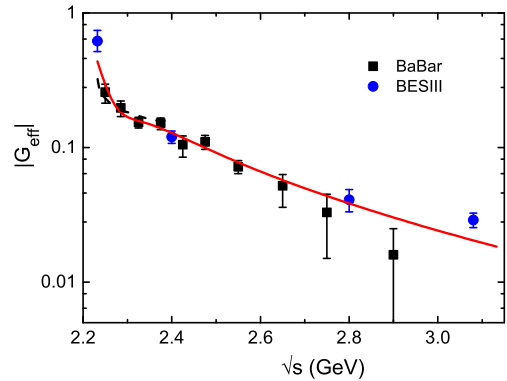}
  \caption{(Color online). Our fit to the effective form factor $|G_{\rm{eff}}|$ of $\Lambda$ hyperon (solid curve). The rectangles and circles represent the data from the BaBar~\cite{Aubert:2007uf} and BESIII~\cite{Ablikim:2017tys,Ablikim:2017pyl} Collaborations, respectively. For comparison, we also present the result from the $\Lambda\bar\Lambda$ potential model in Ref.~\cite{Haidenbauer:2016won} (dashed curve).}\label{geff}
\end{figure}
\begin{figure}[htb]
  \centering
  \includegraphics[width=0.95\columnwidth]{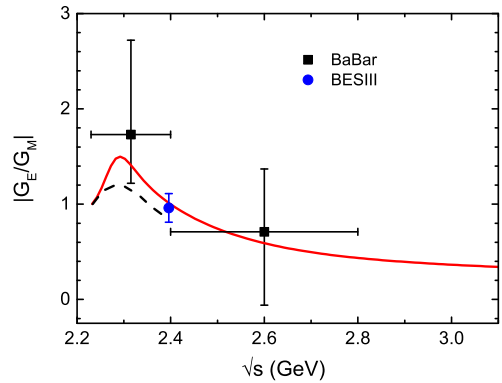}
  \caption{(Color online). The same as Fig. \ref{geff} but for the electromagnetic form factor ratio. The experimental data on the ratio $|G_E/G_M|$ are from BaBar~\cite{Aubert:2007uf}  and BESIII~\cite{Cui:2018} Collaborations.}\label{ratio}
\end{figure}

\begin{figure}[htb]
\centering
  \includegraphics[width=0.95\columnwidth]{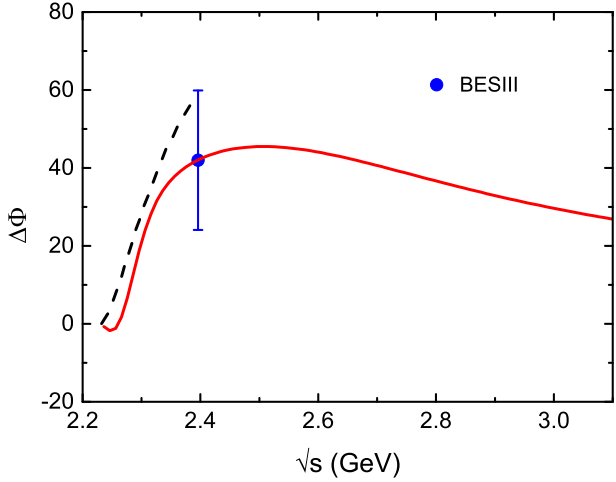}
  \caption{(Color online). The same as Fig. \ref{geff} but for the relative phase $\Delta \Phi$. The experimental data are from BESIII Collaboration~\cite{Cui:2018}.}\label{arg}
\end{figure}

\section{Numerical results and discussions \label{Sec:3}}

\subsection{Fit the time-like form factors}

The masses and widths of the involved vector mesons are listed in Table \ref{Tab:Mass}.  With the expressions of $F_1$ and $F_2$ in Eqs. (\ref{F1s})-(\ref{F2s}) and the replacements in Eq. \ref{Eq:Rep}, we perform a combined fit to the effective form factor $G_{\rm eff}$ in the region $2.2324~\textrm{GeV} < \sqrt{s} <3.08~\textrm{GeV}$, the electromagnetic form factor ratio and the relative phase. The obtained parameters are presented in Table \ref{Tab:Para}. With these parameters, the $\chi^2/\rm{d.o.f}$ is estimated to be 0.977.

\begin{table*}[htb]
\caption{The parameters obtained from the combined fit. \label{Tab:Para}}
\begin{tabular}{p{1.5cm}<{\centering} p{2.2cm}<{\centering} p{1.5cm}<{\centering} p{2.2cm}<{\centering} p{1.5cm}<{\centering} p{2.2cm}<{\centering}}
\toprule[1pt]
Parameter & Value & Parameter &Value & Parameter &Value \\
\midrule[1pt]
$\beta_{\omega(782)}$ & $3.141 \pm 0.004$  &
$\beta_{\omega(1420)}$ & $-0.632 \pm 0.003$  &
$\beta_{\omega(1650)}$ & $0.422\pm 0.003$ \\
$\beta_{\phi(1020)}$ &  $-3.343 \pm 0.004$ &
$\beta_{\phi(1680)}$ & $0.388\pm 0.002$  &
$\beta_{\phi(2170)}$ & $ -0.089 \pm 0.001$\\
$\alpha_{\phi(1020)}$ & $7.079\pm 0.044$  &
$\alpha_{\phi(1680)}$ & $0.910 \pm 0.001$  &
$\alpha_{\phi(2170)}$ & $0.073 \pm 0.001$ \\
\bottomrule[1pt]
\end{tabular}
\end{table*}

In Fig.~\ref{geff}, we present our best fit to the $\Lambda$ effective form factor $|G_{\rm{eff}}|$ in the energy range $2.2324~\textrm{GeV} < \sqrt{s} <3.08~\textrm{GeV}$. For comparison, we also present the theoretical estimations from the potential model up to $\sqrt{s}=2.4 \ \mathrm{GeV}$ \cite{Haidenbauer:2016won}. After including the resonances of $\omega$ and $\phi$ mesons, the fit demonstrates that our model can accurately describe the effective form factor $|G_{\rm{eff}}|$ of the Lambda hyperon. In particular, after including $\phi(2170)$, the near threshold enhancement can be well reproduced.

In Fig .\ref{ratio}, we present the fit result of the electromagnetic form factor ratio. This ratio is determined to be $1$ due to the kinematical restriction on the threshold. With $\sqrt{s}$ increasing, this ratio increased at first and then decreased, which is similar to estimation of the $\Lambda \bar{\Lambda}$ potential model  \cite{Haidenbauer:2016won}. Our fitted results  reach up its maximum $\sim 1.5$ at $\sqrt{s}\simeq 2.3\ \mathrm{ GeV}$, and then decreases  monotonically. While the maximum of the $\Lambda \bar{\Lambda}$ potential model estimation is  about 1.2. Our fit to the relative phase $\Delta \Phi$ is presented in Fig. \ref{arg}. It should be notice that $G_E=G_M$ on the $\Lambda \bar{\Lambda}$ threshold, thus, $\Delta \Phi=0$ on the threshold. Our result well fit the experimental data. In the combined fit, we find that the inclusion of the vector resonance states is crucial for describing the data on $|G_{\rm{eff}}|$, $|G_E/G_M|$ and $\Delta\Phi$. After including these vector resonances, the near threshold enhancements of effective form factor, the large electromagnetic form factor ratio and large phase angle can be well reproduced.

\begin{figure}
  \centering
  \includegraphics[width=0.95\columnwidth]{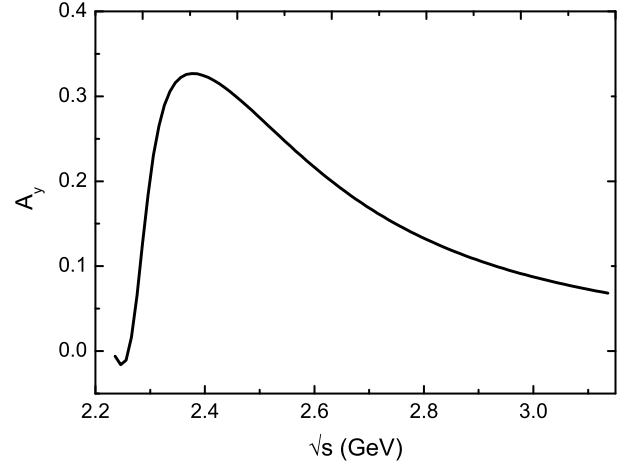}
  \caption{Prediction for the single polarization observable $A_y$ vs $\sqrt{s}$ in $e^+ e^- \rightarrow \Lambda \bar{\Lambda}$ at the fixed angle $\theta=45^\circ$.}
  \label{singlepolar}
\end{figure}

\begin{figure*}
  \centering
  \includegraphics[width=0.85\columnwidth]{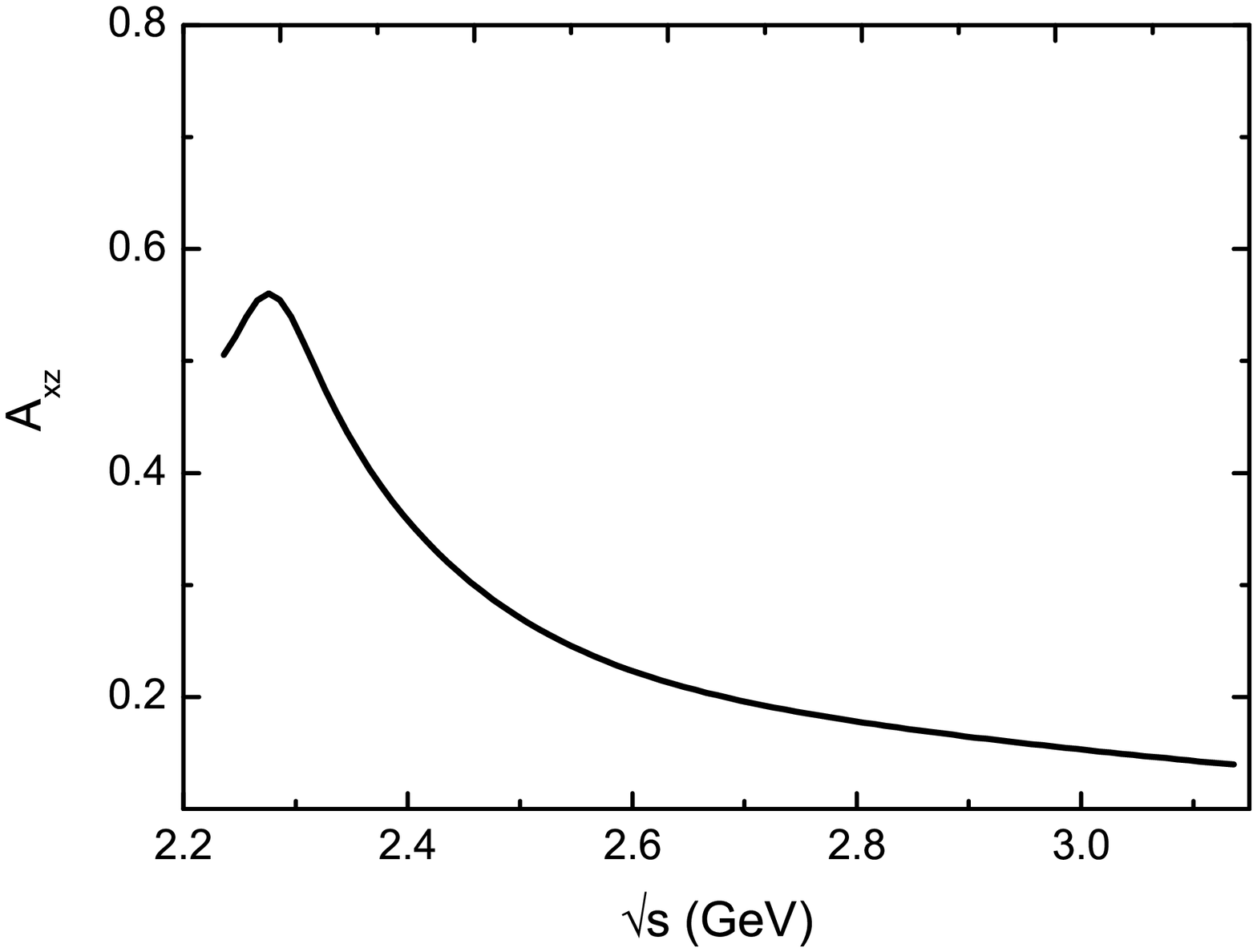}~~~~~~
  \includegraphics[width=0.85\columnwidth]{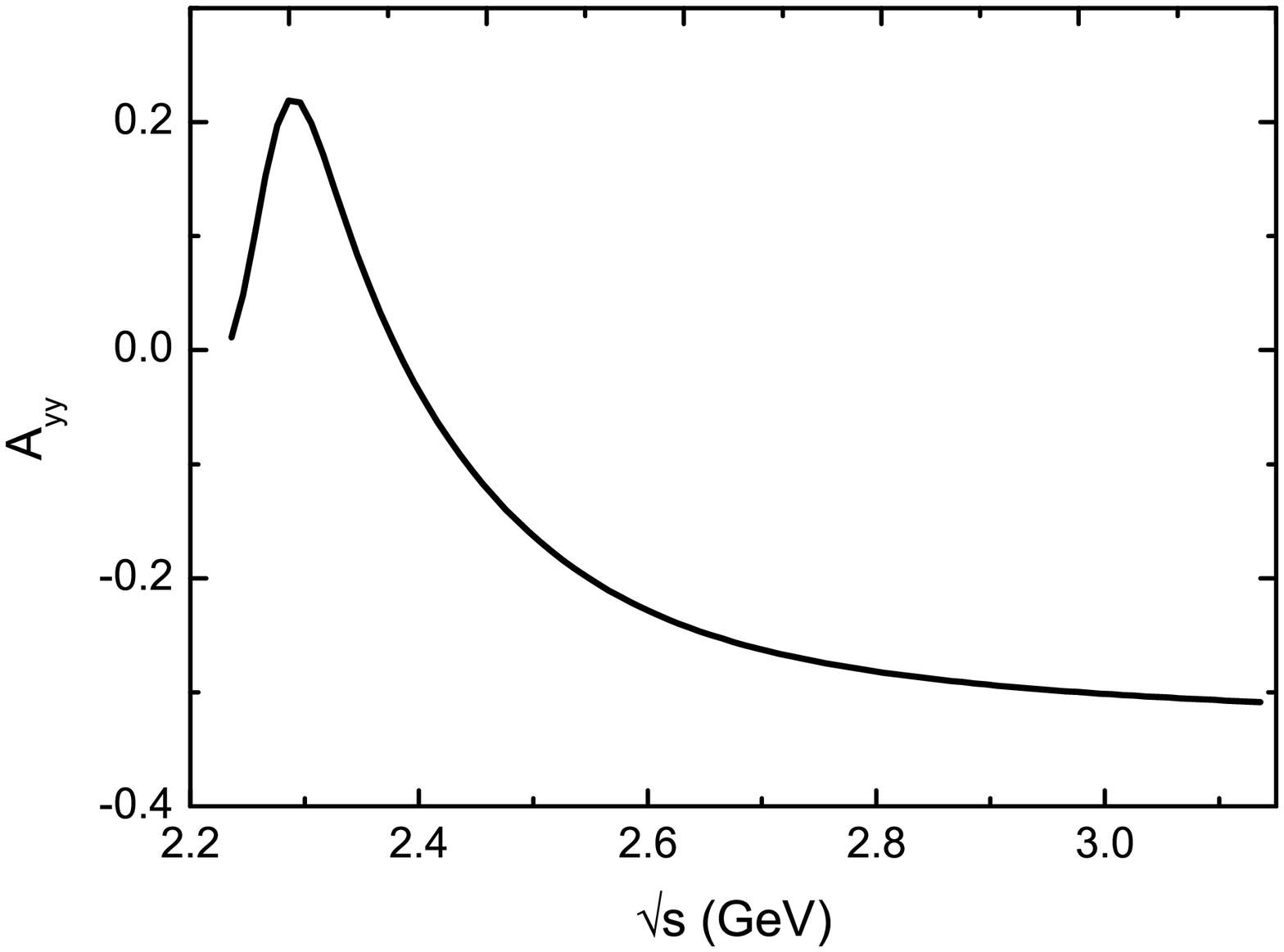}\\
  \includegraphics[width=0.85\columnwidth]{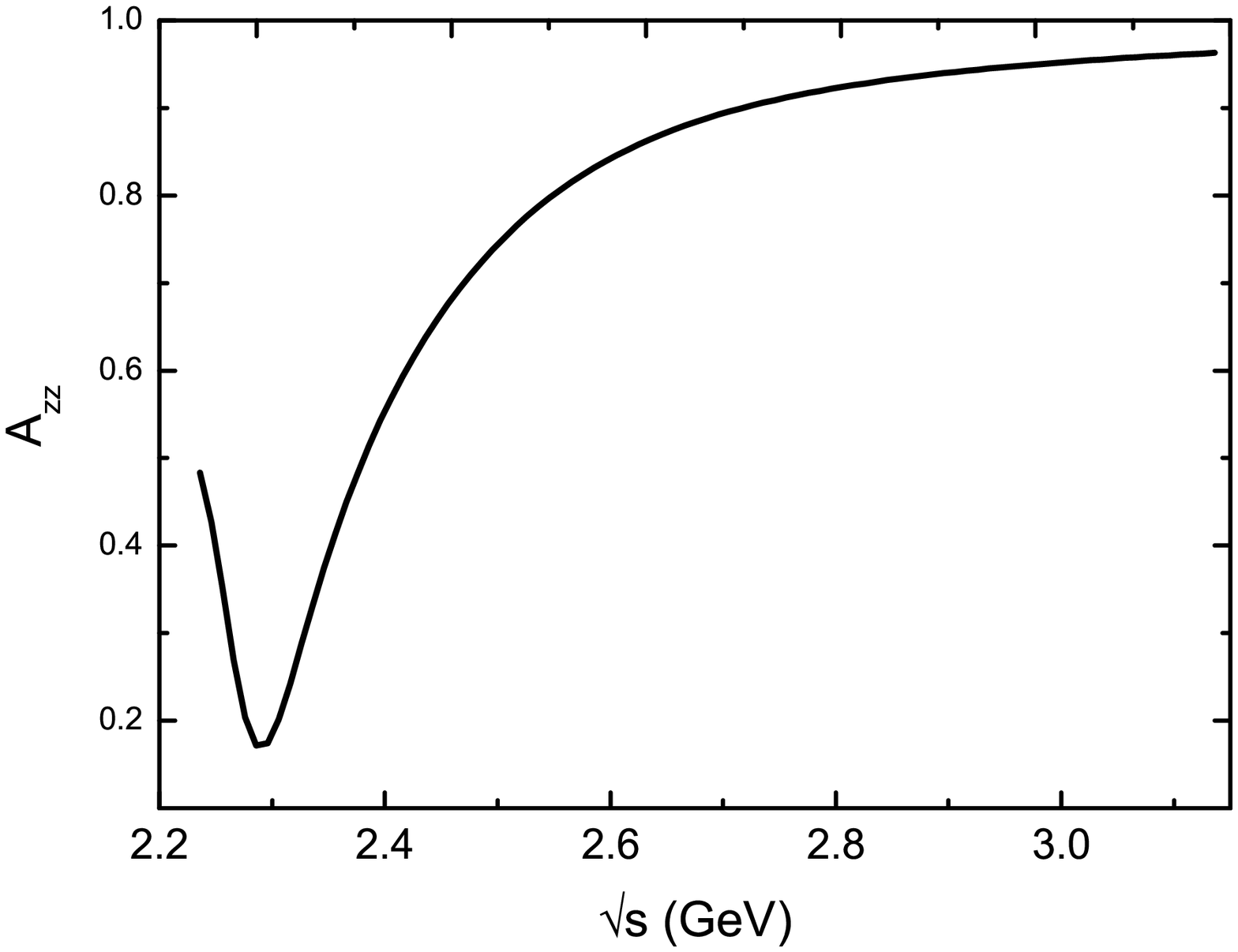}~~~~~~
  \includegraphics[width=0.85\columnwidth]{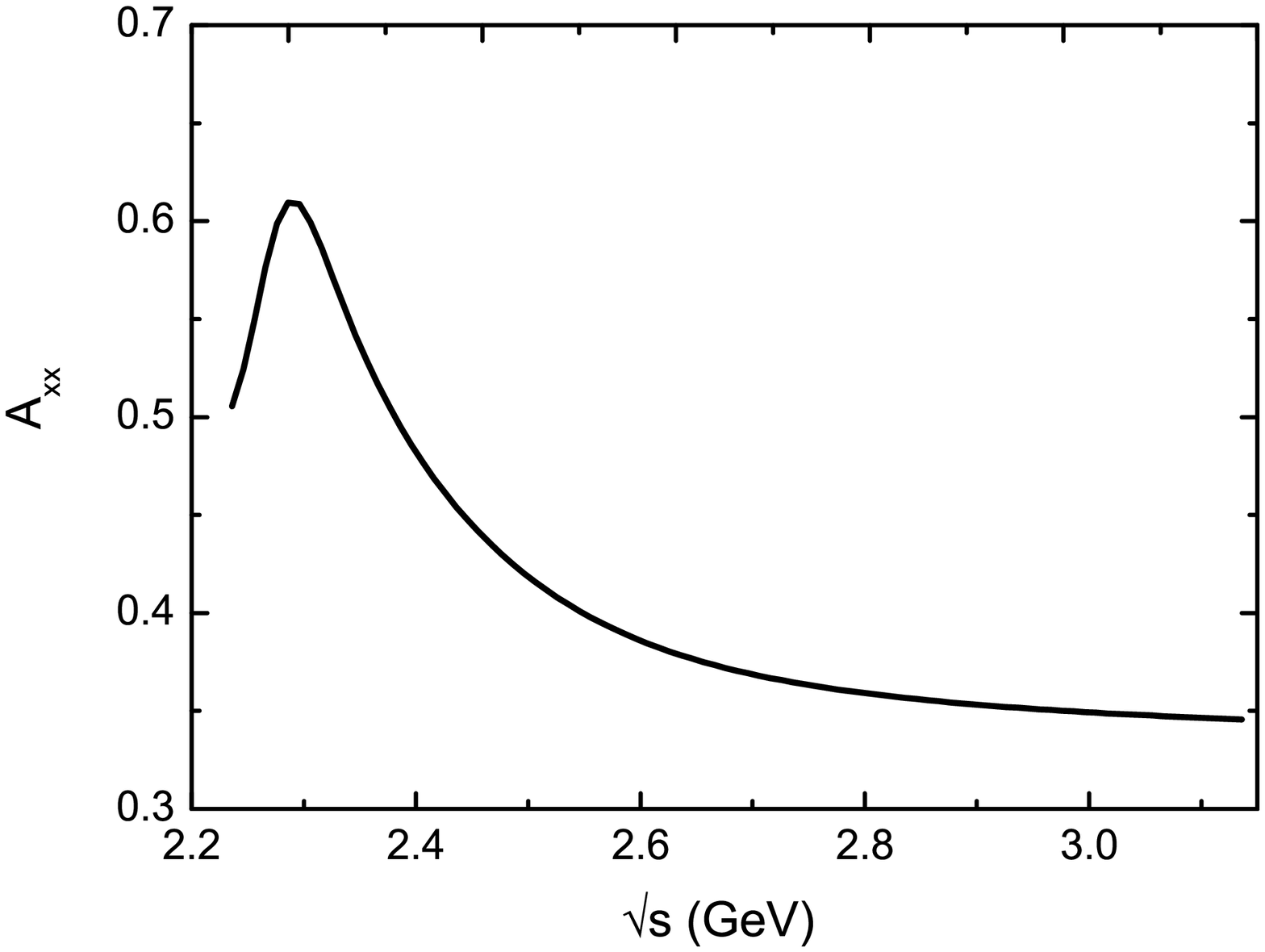}
  \caption{Predictions for the double polarization observables $A_{xz}$, $A_{xx}$, $A_{yy}$ and $A_{zz}$ vs $\sqrt{s}$ in $e^+ e^- \rightarrow \Lambda \bar{\Lambda}$ at the fixed angle $\theta=45^\circ$.}
  \label{doulbepolar}
\end{figure*}

\subsection{Polarization observables in time-like region}

With the fitted $G_E$ and $G_M$, we can estimate the spin polarized observables appearing in the reaction $e^+ e^- \to \bar{\Lambda}\Lambda~$\cite{Faldt:2016qee, Faldt:2017kgy}. Under one-photon exchange approximation, the single and double polarization observables can be related to electromagnetic form factors by,
\begin{eqnarray}
A_{y}&=&{-2M_\Lambda\sqrt{s}\sin(2\theta)~\rm{Im}(G_MG_E^*)\over{D_c-D_s\sin^2(\theta)}}
\,, \nonumber \\
A_{xz}&=&{4M_\Lambda\sqrt{s}\cos(\theta)\rm{Re}(G_MG_E^*)\over{D_c-D_s\sin^2(\theta)}}
\,, \nonumber  \\
A_{xx}&=&{[D_c-D_s]\sin^2(\theta)\over{D_c-D_s\sin^2(\theta)}}\,, \nonumber  \\
A_{yy}&=&{-D_s\sin^2(\theta)\over{D_c-D_s\sin^2(\theta)}}\,, \nonumber \\
A_{zz}&=&{[D_s\sin^2(\theta)+D_c\cos^2(\theta)]\over{D_c-D_s\sin^2(\theta)}}\,,
\label{eq:doublespin3}
\end{eqnarray}
where $D_c=2s|G_M|^2$, $D_s=s|G_M|^2-4M^2|G_E|^2$, and $\theta$ is the scattering angle defined in the c.m. frame. In this work, we present the polarized observables with  $\theta=45^\circ$. From above expressions of the spin polarization observables, one can find that the observables $A_{xx}$, $A_{yy}$ and $A_{zz}$ depend on moduli of the EMFFs only. As for $A_y$ and $A_{xz}$, they
not only depend on the moduli of the EMFFs, but also on the relative phase,
since $G_E/G_M=|G_E/G_M| e^{i \Delta\Phi}$~\cite{Faldt:2017kgy}.
Therefore, precise measurements on these observables will be very useful for discriminating different models.
In Fig.~\ref{singlepolar}, we present the single polarization observable $A_y$ as function of $\sqrt{s}$. Since $\rm{Im}(G_M G_E^*) \sim \sin(\Delta\Phi)$, the shape of $A_y$ can be directly related to the relative phase $\Delta \Phi$, which indicates the precise measurement of the single spin polarization $A_y$ can provide a way to obtain the exact information of the relative phase $\Delta\Phi$. The double polarization observables depending on $\sqrt{s}$ are presented in Fig.~\ref{doulbepolar}.

\subsection{Form factors in space-like region}
It is straightforward to extend our estimation to the space-like region with the fitting parameters.
Unlike the case of the nucleon, the $\Lambda$ hyperon space-like EMFFs are hard to be measured experimentally and no experimental data is available until now. But there are some theoretical investigations on EMFFs $\Lambda$ hyperon in space-like region. In Ref~\cite{Sanchis-Alepuz:2015fcg}, EMFFs of $\Lambda$ hyperon in space-like region are estimated by using a combination of Dyson-Schwinger equation and Bethe-Salpeter equation, where the former equation is used to calculate  the quark propagator and the latter one is adopted to describe the baryons as bound states of three valence quarks. The EMFFs were also estimated in a relativistic quark model,  where both contributions from the quark core and meson cloud are considered \cite{Ramalho:2011pp}. The Lattice QCD estimated the $\Lambda$ hyperon EMFFs at $Q^2=0.227(2)\ \rm{GeV}^2$ with unphysical pion mass \cite{Boinepalli:2006xd}.

\begin{figure*}
  \centering
  \includegraphics[width=0.95\columnwidth]{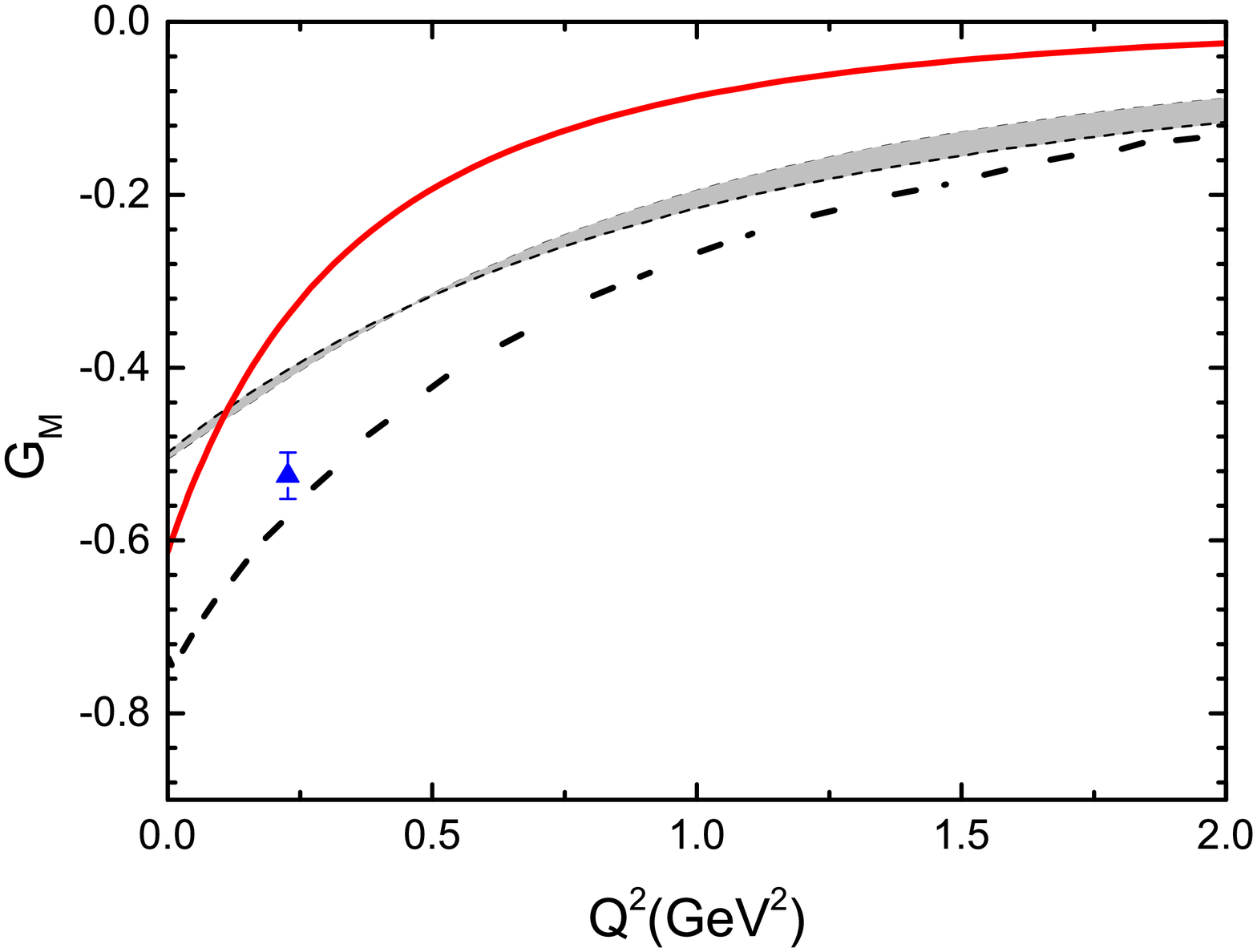}
  \includegraphics[width=0.95\columnwidth]{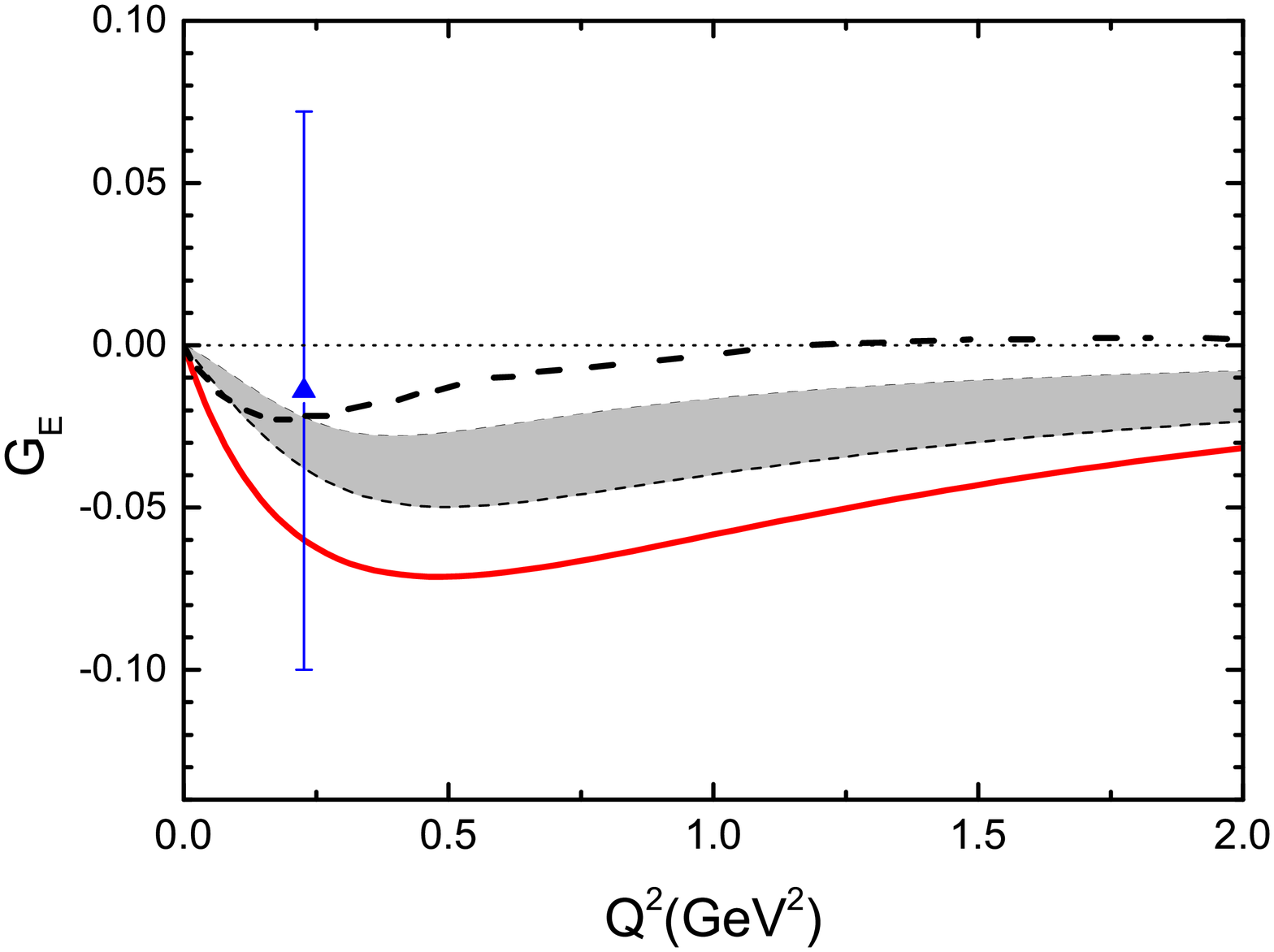}
  \caption{(Color online). The EMFFs of $\Lambda$ hyperon from the present estimation (solid lines) comparing with the predictions of Ref.~\cite{Sanchis-Alepuz:2015fcg} (shadow area), Ref.~\cite{Ramalho:2011pp} (dashed lines) and the Lattice data (triangles point with error bar) from Ref.~\cite{Boinepalli:2006xd}.}
  \label{gmge}
\end{figure*}

With the fitted parameters in Table \ref{Tab:Para}, we present our numerical results of space-like EMFFs in Fig.~\ref{gmge}. For comparison, we also show the estimations of $\Lambda$ hyperon EMFFs from different model ~\cite{Sanchis-Alepuz:2015fcg,Ramalho:2011pp} and Lattice QCD \cite{Boinepalli:2006xd}. Our estimation indicate $G_E$ is negative and the magnitude is larger than that of Refs.~\cite{Sanchis-Alepuz:2015fcg,Ramalho:2011pp}, but the shapes are similar. As for the magnetic form factor, it is also negative and its magnitude decreases more quickly than other model predictions ~\cite{Sanchis-Alepuz:2015fcg,Ramalho:2011pp}.

\section{Summary \label{Sec:4}}

In the present work, we have presented an analysis on the EMFFs of the $\Lambda$ hyperon in a modified VMD model. We take into account the contributions of two components, which are the valence quarks and the meson cloud in the VMD model.  Different from the case of nucleon, the isovector parts of the Dirac and Pauli form factors vanish due to the isoscalar nature of the $\Lambda$ hyperon. We further analytically continue the EMFFs from the space-like region to the time-like region.

To fit the time-like effective form factor $G_{\rm eff}$, electromagnetic form factor ratio $|G_E/G_M|$ and the relative phase angle $\Delta \Phi$ in the time-like region, we involve all the isospin singlet vector meson below the $\Lambda \bar{\Lambda}$ threshold, which are $\omega(782)$, $\omega(1420)$, $\omega(1650)$, $\phi(1020)$, $\phi(1680)$ and $\phi(2170)$. With these vector mesons, we can well reproduce the experimental data, and we also find the excited states $\omega(1420)$, $\omega(1650)$, $\phi(1680)$ and $\phi(2170)$ are crucial in depicting the experimental data.

With the parameters obtained by the combined fit of $|G_{\rm eff}|$, $|G_E/G_M|$ and $\Delta \Phi$, we predicted the single and double spin polarization observables of $e^+e^-\to \Lambda \bar{\Lambda}$ process, which could be tested by further precise measurements. Moreover, we analytically continue the expressions of the form factors from the time-like region to the space-like region. Our estimations are qualitatively similar to other model calculations and Lattice estimation but slight different quantitatively.

\section{Acknowledgements}

This work is partially supported by the NSFC under Grant Nos. 11575043 and 11775050,  by the Fundamental Research Funds for the Central Universities of China.
Y.~Yang is supported by the Scientific Research Foundation of Graduate
School of Southeast University (Grant No. YBJJ1770) and the Postgraduate Research \& Practice Innovation Program of Jiangsu Province (Grants No. KYCX17\_0043).

\end{document}